\documentclass[amssymb,aps,twocolumn,nofootinbib,showpacs]{revtex4}
\usepackage{amsmath}
\usepackage{eufrak}

\def\ket#1{\left| #1 \right\rangle}

\def\bracket#1#2{\left\langle #1 | #2 \right\rangle}
\def\kb#1#2{|#1\rangle\!\langle #2 |}
\def\II{1\!\mathrm{l}}
\def\Tr{\mbox{Tr}}

\newcommand{\Null}{\ensuremath{\mathcal{N}}}

\begin{document}

\newtheorem{lemma}{Lemma}
\newtheorem{definition}{Definition}
\newtheorem{theorem}{Theorem}
\newtheorem{corollary}{Corollary}

\title{Compatibility of quantum states}
\author{David Poulin\footnote{Email address:
    poulinda@iro.umontreal.ca} and
    Robin Blume-Kohout\footnote{Email
    address: rbk@socrates.Berkeley.edu}}
\affiliation{
Los Alamos National Laboratory, Theoretical Division,
MS-B210, Los Alamos, New Mexico 87545}
\date{\today}

\begin{abstract}
We introduce a measure of the compatibility between quantum states---the
likelihood that two density matrices describe the same object. Our measure
is motivated by two elementary requirements, which lead to a natural
definition. 
We list some properties of this measure, and discuss its relation to the
problem of combining two observers' states of knowledge.
\end{abstract}


\maketitle

\date{\today}

The quantum superposition principle induces a qualitative difference
between classical and quantum states of knowledge. The state of a
quantum system can be fully specified, yet not predict with certainty
the outcome of a measurement---a state of affairs which has only the
observers' ignorance as classical analogue. In quantum mechanics,
incomplete knowledge is represented by a mixed density matrix, which
corresponds imperfectly to a classical distribution;
the  ``quantum uncertainty'' of pure states combines with the
``classical uncertainty'' of a distribution to yield an object which
can be represented by different decompositions or preparations.

The fidelity of two quantum states \cite{Jozsa1994} $\rho_A$ and $\rho_B$,
\begin{equation}
F(\rho_A,\rho_B) =
\Tr\left\{\sqrt{\sqrt{\rho_A}\rho_B\sqrt{\rho_A}}\right\}
\label{eq_fidelity}
\end{equation}
(or more precisely $F^2$) measures the likelihood that various
measurements made on the two
states will obtain the same result.  Thus, fidelity is a measure
of similarity between states which does not distinguish between
classical and quantum uncertainty.

In this letter, we introduce \emph{compatibility}, a measure similar
to fidelity, but which compares two observers' states of knowledge,
not the results of the measurements which they could do.  We want the
compatibility to measure classical admixture, while treating different
pure states as fundamentally different: if two observers claim to have
complete knowledge of a system, their descriptions had better agree
completely. Hence, a compatibility measure
$C(\rho_A,\rho_B)$ should satisfy the two following requirements:
\begin{enumerate}
\item When $[\rho_A,\rho_B]=0$ (classical mixture) the compatibility
  should be equal to the fidelity.
\item The compatibility of incompatible states should be~0.
\end{enumerate}

While our first requirement should be transparent, the second sounds
tautological, and requires further explanation.  Consider two
observers (Alice and Bob) whose respective states of knowledge are
described by $\rho_A$ and $\rho_B$. (Throughout this letter, we use
subscript $k$ to designate
either $A$ or $B$.) Brun, Finkelstein, and Mermin
\cite{BFM2002} defined Alice's and Bob's descriptions to be {\it compatible}
if and only if they \emph{could} be describing the same physical
system.  They then addressed the following question: under what
conditions are $\rho_A$ and $\rho_B$ compatible?  Their answer is
quite simple: $\rho_A$ and $\rho_B$ are compatible if and only if the
intersection of their supports, $\mathcal S = \mathcal S(\rho_A)\cap
\mathcal S(\rho_B)$, is nonempty. The support
$\mathcal{S}(\rho)$ of a density matrix $\rho$ is the complement of
its null space $\mathcal N(\rho)$; to obtain the projector
$P_{\mathcal{S}(\rho)}$ onto 
$\mathcal{S}(\rho)$, diagonalize $\rho$ and replace each nonzero
eigenvalue with 1.  Thus, $\rho_A$ and $\rho_B$ are compatible if
$P_{\mathcal{S}(\rho_A)}P_{\mathcal{S}(\rho_B)}$ has at least one unit
eigenvalue. In other words, two states of knowledge are incompatible
if between them they rule out all possible pure states.

With this definition, state $\rho_A=\kb{0}{0}$ is compatible with both
state $\rho_B = \epsilon \kb{0}{0} + (1-\epsilon)\kb{1}{1}$
and state $\rho_B' = (1-\epsilon) \kb{0}{0} + \epsilon\kb{1}{1}$ as
long as $0<\epsilon <1$. Nevertheless as $\epsilon \to 0$, it is clear
that the compatibility of $\rho_A$ and $\rho_B$ should vanish while
that of $\rho_A$ and $\rho_B'$ should approach unity. The
definition of \cite{BFM2002} makes no distinction between these two
cases and this is what originally motivated the present work.

Now that requirement 2 has been clarified, we can proceed with the
definition of the compatibility measure.

\begin{definition}
Let $\mathfrak{B}_0(\mathcal H)$ be the set of all density matrices on
Hilbert space $\mathcal H$. For $\rho \in \mathfrak{B}_0(\mathcal H)$,
define $\mathcal P(\rho)$ as the set of preparations of
$\rho$:
\label{definition}
\begin{equation}
\mathcal P(\rho) =
\left\{ P : \int_{\mathfrak{B}_0(\mathcal H)} P(\sigma)\sigma d\sigma = \rho\right\}
\end{equation}
where the $P$ are probability distributions over $\mathfrak{B}_0(\mathcal
H)$. Then, the compatibility of $\rho_A$ and $\rho_B \in
\mathfrak{B}_0(\mathcal H)$ is \hbox{defined as}
\begin{equation}
C(\rho_A,\rho_B) = \max_{\substack{P_A\in\mathcal P(\rho_A) \\
P_B\in \mathcal P(\rho_B)}}
\int_{\mathfrak{B}_0(\mathcal H)}
\sqrt{P_A(\sigma)P_B(\sigma)}d\sigma,
\label{eq_def_mixe}
\end{equation}
the integral representing the classical fidelity $F(P_A,P_B)$ (or
statistical overlap) of two classical distributions $P_A$ and $P_B$.
\end{definition}

\begin{lemma}
All distributions $P \in \mathcal P(\rho)$ must vanish outside $\mathfrak
B_0(\rho)$: the set of density
matrices with support restricted to $\mathcal S(\rho)$ (this is a
slightly abusive notation).   
\end{lemma}
\noindent{\it Proof}
Let $P(\sigma)$ be a preparation of $\rho$. We can separate $\rho$
in two parts:
\begin{eqnarray}
\rho &=&
\int_{{\mathfrak{B}}^\bot_0(\rho)}P_k(\sigma)\sigma\ d\sigma
+ 
\int_{{\mathfrak{B}}_0(\rho)}P_k(\sigma)\sigma\ d\sigma \\
&=& p\rho' + (1-p)\rho''
\end{eqnarray}
where $\rho'$, by definition, has support on $\mathcal
N(\rho)$, $\rho''$ has support strictly on $\mathcal
S(\rho)$ and $p = 
\int_{{\mathfrak{B}}^\bot_0(\rho)} P_k(\sigma)\ d\sigma$. If
$p \neq 0$, 
there exists $\psi \in\Null(\rho)$ such that $\rho$
does not annihilate $\psi$.  This contradicts the definition of
$\Null(\rho)$, so we conclude that $p = 0$ and therefore $P$ is
restricted to $\mathfrak B_0(\rho)$.

\begin{theorem}
Definition 1 satisfies both of our requirements.
\end{theorem}
\noindent{\it Proof}

1.
  If $\rho_A$ commutes with $\rho_B$, then they have orthogonal
  decompositions onto the same set of pure states:
  $\rho_A=\sum_i{a_i\kb{\phi_i}{\phi_i}}$;
  $\rho_B=\sum_i{b_i\kb{\phi_i}{\phi_i}}$. Thus
  $C(\rho_A,\rho_B)\geq\sum_i{\sqrt{a_ib_i}} = F(\rho_A,\rho_B)$.
  Later (see P4) we show that $C(\rho_A,\rho_B)\leq
  F(\rho_A,\rho_B)$; therefore for commuting density matrices
  $C(\rho_A,\rho_B)=F(\rho_A,\rho_B)$.

2.
  If $\rho_A$ and $\rho_B$ are
  incompatible, their supports are disjoint, which implies that
  $P_A(\sigma)$ and $P_B(\sigma)$ are restricted to disjoint
  sets---implying that $C(\rho_A,\rho_B)=0$. 

Note that this measure is not the only on which satisfies our two
requirements. For example, define
\begin{equation}
D_n(\rho_A,\rho_B) = 
Tr\left\{\left[(\rho_A)^{1/2n}(\rho_B)^{1/n}(\rho_A)^{1/2n}\right]^{n}\right\}.
\label{eq_measure2}
\end{equation}
Clearly, $D_n(\rho_a,\rho_B) = F(\rho_A,\rho_B)$ when $n=1$ or for any
$n$ when $[\rho_A,\rho_B]=0$, so it satisfy our first requirement. For the
second requirement, notice that 
\begin{equation}
\lim_{n \to \infty} D_n(\rho_A,\rho_B) \leq
\lim_{n \to \infty} Tr\left\{\left[
P_{\mathcal S(\rho_A)}P_{\mathcal S(\rho_B)}
\right]^n\right\}
\end{equation} which is $0$ if $P_{\mathcal S(\rho_B)}P_{\mathcal
  S(\rho_B)}$ has no unit eigenvalue, i.e. if $\rho_A$ and $\rho_B$
are incompatible. Therefore $D(\rho_A,\rho_B) = \lim_{n\to\infty}
  D_n(\rho_A,\rho_B)$ is a valid measure 
of compatibility. 

Definition \ref{definition} can also be generalized to 
\begin{equation}
E_\alpha(\rho_A,\rho_B) = \max_{\substack{P_A\in\mathcal P(\rho_A)\\
P_B\in\mathcal P(\rho_B)}}
\int_{\mathfrak B_0(\mathcal H)} [P_A(\sigma)]^\alpha[P_B(\sigma)]^{1-\alpha}d\sigma 
\label{eq_def_E}
\end{equation}
$0< \alpha < 1$
which is the R\'enyi overlap of $P_A$ and $P_B$, the fidelity
corresponding to the
special case $\alpha = 1/2$. This definition allows for an asymmetry
between Alice and Bob which can be useful when one of the participant
is more trustworthy than the other.

Although these alternative definitions offers some interesting
features, we shall concentrate on Definition \ref{definition} in the
following. (Superscript $D$ and $E$ indicates that the results also hold for
measure $D(\rho_A,\rho_B)$ and $E_\alpha(\rho_A,\rho_B)$ respectively,
the proofs are given for $C(\rho_A,\rho_B)$ only.)

\begin{theorem}$^E$
To compute the compatibility of two states, it is sufficient to
maximize over pure state preparations. In other words
\begin{equation}
C(\rho_A,\rho_B) = \max_{\substack{Q_A\in\mathcal Q(\rho_A) \\
Q_B\in\mathcal Q(\rho_B)}}
\int_{\mathfrak B_0^1(\mathcal H)} \sqrt{Q_A(\psi)Q_B(\psi)}d\psi
\label{eq_def_pure}
\end{equation}
where $\mathfrak B_0^1(\mathcal H)$ is the set of all pure states in $\mathcal H$ and
$\mathcal Q(\rho)$ is the set of pure state preparations of $\rho$:
\begin{equation}
\mathcal Q(\rho) =
\left\{ Q : \int_{\mathfrak B_0^1(\mathcal H)} Q(\psi)\kb{\psi}{\psi} d\psi =
  \rho\right\},
\end{equation}
$Q$ are probabilities distributions on $\mathfrak B_0^1(\mathcal H)$.
\label{th_pure}
\end{theorem}
\noindent{\it Proof} Choose a standard preparation for $\sigma \in
  \mathfrak B_0(\mathcal H)$,
$\sigma = \int_{\mathfrak B_0^1(\mathcal H)}f_\sigma (\psi) \kb{\psi}{\psi}
  d\psi$ (e.g., eigendecomposition). Then
\begin{eqnarray}
&&\int_{\mathfrak B_0(\mathcal H)} \sqrt{P_A(\sigma)P_B(\sigma)}\
d\sigma \\
&=&\int_{\mathfrak B_0(\mathcal H)}\int_{\mathfrak B_0^1(\mathcal H)}
\sqrt{P_A(\sigma)f_\sigma(\psi)P_B(\sigma)f_\sigma(\psi)}\ d\sigma\
d\psi \nonumber \\
&\leq& \int_{\mathfrak B_0^1(\mathcal H)}\sqrt{Q_A(\psi)Q_B(\psi)}\ d\psi \nonumber
\end{eqnarray}
since fidelity can only increase under the marginalization $Q_k(\psi)
  = \int_{\mathfrak B_0(\mathcal H)} P_k(\sigma)f_\sigma(\psi)\ d\sigma$.

\begin{theorem}
When one of the two states is pure (say $\rho_B$),
$C(\rho_A,\rho_B) = \sqrt{p}$ where $p$ is given by
\begin{equation}
p = \min_{q \in [0,1]}\left\{q: \det_{\mathcal S(\rho_A)}
\{\rho_A-q\rho_B\} = 0\right\}
\label{eq_q}
\end{equation}
if $\rho_B$ lies within $\mathcal S(\rho_A)$ and $p=0$ otherwise.
\end{theorem}
\noindent{\it Proof}
There is a unique preparation for $\rho_B$:
$P_B(\sigma)=\delta(\sigma-\rho_B)$. 
The maximum value of $q$ for which we can write $\rho_A = q\rho_B +
  (1-q)\sigma$ 
(with $\sigma$ a valid density matrix) is $p$. The result follows. 

\begin{theorem}$^E$
Any local maximum of $F(P_A,P_B)$ over $\mathcal P(\rho_A) \otimes
  \mathcal P(\rho_B)$ is a global maximum.
%Furthermore, the maximum is
%attained on external points of the set.
\end{theorem}
\noindent{\it Proof}
Fidelity is a concave function: $F(\lambda P_A +$
\hbox{$[1-\lambda] P_A',P_B) \geq \lambda F(P_A,P_B) +
  [1-\lambda] F(P_A',P_B)$}. The sets $\mathcal P(\rho_A)$ and
$\mathcal P(\rho_B)$ are convex: any convex combinations of valid
probability distributions of mean $\rho$ is also a valid probability
distribution with mean $\rho$. The result follows
automatically.

We now give a list of properties of the compatibility measure.
\medskip

\noindent{\bf P1}$^D$
$C(\rho_A,\rho_B)$ is symmetric.
\medskip

\noindent{\bf P2}$^{DE}$
Compatibility is invariant under unitary transformation:
$C(U\rho_AU^\dagger,U\rho_BU^\dagger) = C(\rho_A,\rho_B)$.
\medskip

\noindent{\bf P3}$^{DE}$
For pure states $C(\psi_A,\psi_B) = 1$ if and only if
$|\bracket{\psi_A}{\psi_B}|^2 =1$ and 0 otherwise.
\medskip

\noindent{\bf P4}$^D$
(Upper bound) $C(\rho_A,\rho_C) \leq F(\rho_A,\rho_B)$.
\medskip

\noindent{\bf P5}$^{DE}$
$F(\rho_A,\rho_B)=0 \Rightarrow C(\rho_A,\rho_B) =0$ and
$F(\rho_A,\rho_B)=1 \Leftrightarrow C(\rho_A,\rho_B) =1
\Leftrightarrow \rho_A = \rho_B$.
\medskip

\noindent{\bf P6}
(Lower bound) $C(\rho_A,\rho_B) \geq
r\sqrt{\epsilon_A\epsilon_B}$ 
where $\epsilon_k$ is the greatest value of $q$ for which one can
write $\rho_k = \frac{q}{r}P_\mathcal S + (1-q)\sigma$ with $\sigma$
being a valid density matrix, see (eq.\ref{eq_q}),
and $r=Tr\{P_\mathcal S\}$ is the dimension of $\mathcal S
= \mathcal S(\rho_A)\cap\mathcal S(\rho_B)$. (For compatible states,
$\epsilon_k \geq \lambda_k^0$, the smallest nonzero eigenvalue of $\rho_k$.)
\medskip

\noindent{\bf P7}$^E$
(Multiplicativity) $C(\rho_A\otimes\rho'_A,\rho_B\otimes\rho'_B) \geq
C(\rho_A,\rho_B)C(\rho'_A,\rho'_B)$.
\medskip

\noindent{\it Proofs}
\medskip

\noindent P1, P2, and P3 are straightforward from Definition
\ref{definition}.
\medskip

\noindent P4: Assume that $Q_{k}(\psi)$ are the optimal
distributions given by Theorem \ref{th_pure}. We choose
$\vec x \in \mathbb R^{(2N-2)}$ (where $N$ is the dimension of
$\mathcal H$)  as a
parameterization for $\mathfrak B_0^1(\mathcal H)$: 
$\psi = \psi(\vec{x})$,
and construct the purifications
\begin{equation}
\ket{\Psi_{k}} = \int \sqrt{Q_{k}(\psi(\vec{x}))}
\ket{\psi(\vec{x})}\otimes\ket{\vec x}\ d\vec x
\label{pur_comp}
\end{equation}
where $\vec x$ is now treated as a quantum continuous variable
$\bracket{\vec x}{\vec x'} = \delta(\vec{x}
-\vec{x}')$ (e.g. position of a particle in a $N$-dimensional box). Then
$C(\rho_A,\rho_B) = \bracket{\Psi_A}{\Psi_B} \leq
F(\rho_A,\rho_B)$ since the fidelity is the maximum
of this quantity over all purifications.

This proof introduces an interesting distinction between fidelity
and compatibility. Fidelity is the optimal inner product between all
purifications of $\rho_A$ and $\rho_B$. On the other hand,
compatibility involves purifications of a very special kind
(eq.\ref{pur_comp}). All that
is needed to transform compatibility into fidelity is to replace
(eq.\ref{pur_comp}) by
\begin{equation}
\begin{array}{c}
\ket{\Psi_{A}} = \int \sqrt{Q_{A}(\psi(\vec{x}))}
\ket{\psi(\vec{x})}\otimes U_A\ket{\vec x}\ d\vec x \\
\ket{\Psi_{B}} = \int \sqrt{Q_{B}(\psi(\vec{x}))}
\ket{\psi(\vec{x})}\otimes U_B\ket{\vec x}\ d\vec x
\end{array}
\end{equation}
for arbitrary unitary operators $U_A$ and $U_B$.
\medskip

\noindent P5 follows from $F(\rho_A,\rho_B) = 1 \Leftrightarrow
\rho_A=\rho_B$, requirement 1, and P4.
\medskip

\noindent P6: We can choose a distribution where $\rho_k$ has
probability $r\epsilon_k$ at the point $\sigma = P_{\mathcal S} /r$.
\medskip

\noindent P7: The product of the optimal distributions for
$C(\rho_A,\rho_B)$ and $C(\rho'_A,\rho'_B)$ are valid distributions
over the combined Hilbert space but might not be optimal. We do not
know if this inequality can be reduced to an equality. In other words,
it is possible that the optimal distribution for
$\rho_A\otimes\rho'_A$ and $\rho_B\otimes\rho'_B$ involve non product
states.

\medskip

It is worth mentioning that no smooth function of the compatibility
satisfying $f(C)=1 \Leftrightarrow C=0$ and $f(C)=0 \Leftrightarrow
C=1$ can be used to build a metric on $\mathfrak B_0(\mathcal H)$. This is best
illustrated by the following 2-dimensional example. Assume states $\rho_+$
and $\rho_-$ are pure, derived from $\ket{\psi_\pm} = \cos\epsilon\ket{0} \pm
  \sin\epsilon\ket{1}$, and $\rho_0 = (1-\epsilon)\kb{0}{0} +
  \epsilon\kb{1}{1}$ where $\epsilon \to 0$. One can easily verify
that $C(\rho_+,\rho_-) = 0$ and $C(\rho_{+},\rho_0) = C(\rho_-,\rho_0)
  = 1 -O(\epsilon)$ so $f(C(\rho_+,\rho_-)) = 1 > f(C(\rho_+,\rho_0)) +
  f(C(\rho_-,\rho_0)) \to 0$ as $\epsilon\to 0$. This is in contrast
with classical distributions: when $[\rho_A,\rho_B]=0$,
$\cos^{-1}F(\sqrt{\rho_A\rho_B})$ is a valid distance measure~\cite{Fuchs1995}.  

\medskip

\noindent
{\it Measurement}---Suppose that Alice and Bob acquire their
knowledge of $\rho_A$ and $\rho_B$ through measurement. These states will
always be compatible: incompatible knowledge acquired through
measurement would indicate an inconsistency in quantum
theory\footnote{Incompatible knowledge could emerge as a consequence
  of the finite accuracy of the measurement apparatus: nevertheless,
  such limitations should be taken into account in the state estimation.}. For
example, they can each be given many copies of a quantum system in
state $\rho$ of which they initially
have no knowledge except the dimension. They carry out independent
measurements on 
those copies and, with the help of Bayesian rules, update their
description of the system (see \cite{SBC2001} and references
therein). As mentioned earlier, their descriptions
will always be compatible.
Nevertheless, a low compatibility could result as a
consequence of one of the following situations: $i$)~they were given
copies of different states; $ii$)~their measurement apparatus are
miscalibrated; or $iii$)~they are in a very improbable branch of the
Universe.

These eventualities cannot be detected by the fidelity of $\rho_A$ and
$\rho_B$. For example, suppose that, for a 2-level system,
\begin{equation}
\begin{array}{c}
\rho_A =(1-\epsilon)\kb{0}{0} + \frac{\epsilon}{2}\II \\
\rho_B =(1-\epsilon)\kb{+}{+} + \frac{\epsilon}{2}\II, \\
\end{array}
\end{equation}
where $\ket + = \frac{1}{\sqrt 2}(\ket 0 + \ket 1)$. 
As the observers'
knowledge becomes more and more accurate ($\epsilon \to 0$), the
compatibility goes to $0$, indicating one of the three situations
listed above. On the other hand, fidelity saturates at $F^2 = 1/2$,
which is the same as if both Alice and Bob had a vague knowledge of the state,
e.g.
\begin{equation}
\begin{array}{c}
\rho_A =(\frac 12 +a)\kb{0}{0} + (\frac 12-a)\kb{1}{1} \\
\rho_B =(\frac 12 -a)\kb{0}{0} + (\frac 12 +a)\kb{1}{1} \\
\end{array}
\end{equation}
with $a = \sqrt{2}/4$. This clearly illustrates the fact that fidelity
makes no distinction between classical and quantum uncertainty.

\medskip

\noindent
{\it Combining knowledge}---Now, suppose Alice and Bob want to pool their
information. If $C(\rho_A,\rho_B)=0$ (which cannot result from
measurement), their ``knowledge'' is
contradictory. When $C(\rho_A,\rho_B) > 0$, however, they can combine
their states of knowledge to get a new density matrix $\rho_{AB}$. This
issue has recently been studied by Jacobs \cite{Jacobs2002} but with
the only conclusion that $\rho_{AB}$ should lie in $\mathcal S(\rho_A) \cap
  \mathcal S(\rho_B)$.

We propose that the state obtained from combining two states of
knowledge should be the one which is maximally compatible with both of
them. This requires a definition of three-way compatibility:
\begin{eqnarray}
&&C(\rho_A,\rho_B,\rho_C) = \\
&&\max_{\substack{P_A\in\mathcal P(\rho_A)\\
P_B\in\mathcal P(\rho_B)\\
P_C\in\mathcal P(\rho_C)}}
\int_{\mathfrak B_0(\mathcal H)}
\sqrt[3]{P_A(\sigma)P_B(\sigma)P_C(\sigma)}d\sigma.
\label{def_3way}
\nonumber
\end{eqnarray}
Hence, our rule for combining states of knowledge reads
\begin{equation}
\rho_{AB} = \textrm{Argument}\left(\max_\rho C(\rho_A,\rho_B,\rho)\right);
\end{equation}
in the
eventuality that the maximum over $\rho$ is not unique, one can
discriminate with a maximum entropy $S(\rho)$ criteria which is well
motivated in the current context.
For any fixed $P_A$ and $P_B$, the $P_C$ that optimizes
(eq.\ref{def_3way})
is proportional to the geometric average of $P_A$ and $P_B$.
Therefore, defining $\tilde P_A$ and $\tilde P_B$ as the distributions
which optimized equation (\ref{eq_def_mixe}), we get
\begin{equation}
\rho_{AB} = \int_{\mathfrak B_0(\mathcal H)} P_{AB}(\sigma)\sigma\ d\sigma
\end{equation}
where $P_{AB} = \sqrt{P_AP_B}/C(\rho_A,\rho_B)$.
Furthermore, there is a simple relation between the optimal three-way
compatibility and the compatibility of the two original descriptions:
$C(\rho_A,\rho_B,\rho_{AB})^3 = C(\rho_A,\rho_B)^2$.

\medskip
\noindent{\it Knowledge}---Knowledge of a quantum system can take many
forms; as Bennett expresses it, 
\begin{quote}
It is possible to {\it know}
or {\it possess} a quantum state in infinitely many physically
inequivalent ways, ranging from complete classical knowledge, through
possession of a single specimen of the state, to weaker and less
compactly embodiable forms such as the ability to simulate the outcome
of a single POVM measurement on the state. \cite{Bennett2001}
\end{quote} 

The compatibility measurement of (eq.\ref{eq_def_mixe}) is meaningful when we consider
classical description of the quantum states; the quantum
fidelity (eq.\ref{eq_fidelity}) corresponds to a situation where single
specimens of the quantum states are available (respectively
``knowledge of the quantum'' and ``quantum knowledge''). One can
define compatibility 
measurements according to the {\it type of knowledge} one is dealing
with. For example, we can define the compatibility between a state
$\rho$ and an ensemble $\{ q_j, \sigma_j\}$ as $\max_{\mathcal P(\rho)}
F(P,Q)$ where $Q(\sigma) = \sum_j q_j\delta(\sigma-\sigma_j)$. While
the pure state $\ket{+} = \frac{1}{\sqrt 2}(\ket{0}+\ket{1})$ is
compatible with the ensemble $E_1 = \{(1,p\kb 00 + (1-p)\kb 11)\}$, it is
incompatible with the ensemble $E_2 = \{(p, \kb 00),((1-p),\kb 11)\}$,
even if they are preparations of the same state. 

An ensemble embodies more knowledge than its associated (average) state. In our
prescription for combining knowledge, we have assumed that all of
Alice's and Bob's knowledge was encapsulated in their respective density
matrices. Note that all knowledge can be represented in this form by
including ancillary systems [e.g. eq.(\ref{pur_comp})].

Suppose, instead,
that both Alice's and Bob's states of knowledge are represented by
the ensemble $E_1$. Obviously, their combined density matrix should be
$\rho_{AB1} = p\kb{0}{0} +$\mbox{$(1-p)\kb{1}{1}$}. On the other
hand, when both their states of knowledge are $E_2$,
Bayesian rules would suggest that their combined state should be
$\rho_{AB2} = p^2\kb{0}{0} +$ \mbox{$(1-p)^2\kb{1}{1}$} (with proper
normalization)---but this assumes that their knowledge was acquired {\it
  independently} \cite{Jacobs2002}.  If their knowledge came from a redundant source,
the Bayesian rule would then yield state $\rho_{AB1}$, as would our
prescription.

Hence, this illustrates that our rule for combining states of knowledge
assumes no more information than what is encapsulated in the density
matrices. Furthermore, it can quite simply be adapted to different
forms of knowledge, either through the use of ancillary systems or of
generalized compatibility measures.

\medskip
\noindent {\it Acknowledgement}s---The authors would like to
acknowledge Howard Barnum, Kurt Jacobs, Harold Ollivier, and Wojciech
Zurek for
discussions on this subject.  RBK would also like to thank Todd Brun
for discussions and inspiration.

\end{document}